\documentclass[prb,twocolumn,showpacs,preprintnumbers,amsmath,amssymb]{revtex4}
\usepackage{graphicx}% Include figure files
\usepackage{dcolumn}% Align table columns on decimal point
\usepackage{bm}% bold math

\begin{document}

\preprint{submitted to Phys. Rev. B}

\title{The origin of paramagnetic magnetization in field-cooled YBa$_2$Cu$_3$O$_{7-\delta}$ films}

\author{D. A. Luzhbin$^1$}
\author{A. V. Pan$^2$}
\email{pan@uow.edu.au}
\author{V.A. Komashko$^1$}
\author{V. S. Flis$^1$}
\author{V. M. Pan$^1$}
\author{S. X. Dou$^2$}
\author{P. Esquinazi$^3$}

\affiliation{$^1$Department of Superconductivity, Institute for Metal Physics, Vernadsky Boulevard 36, Kiev 03142, Ukraine \\
$^2$Institute for Superconducting and Electronic Materials, University of Wollongong, \\ Northfields Avenue, Wollongong, NSW 2522, Australia \\
$^3$Department of Superconductivity and Magnetism, Institute for Experimental Physics II, \\ University of Leipzig, Linn\`estra{\ss}e 5, D-04103 Leipzig, Germany}

\date{June 25, 2003}

\begin{abstract}
Temperature dependences of the magnetic moment have been measured in YBa$_2$Cu$_3$O$_{7-\delta}$ thin films over a wide magnetic field range ($5 \le H \le 10^4$~Oe). In these films a paramagnetic signal known as the paramagnetic Meissner effect has been observed. The experimental data in the films, which have strong pinning and high critical current densities ($J_c \sim 2 \times 10^6$~A/cm$^2$ at 77~K), are quantitatively shown to be highly consistent with the theoretical model proposed by Koshelev and Larkin [Phys. Rev. B {\bf 52}, 13559 (1995)]. This finding indicates that the origin of the paramagnetic effect is ultimately associated with nucleation and inhomogeneous spatial redistribution of magnetic vortices in a sample which is cooled down in a magnetic field. It is also shown that the distribution of vortices is extremely sensitive to the interplay of film properties and the real experimental conditions of the measurements.

\end{abstract}


\maketitle

\section{Introduction}
\label{int}

One of the fundamental properties of superconductors is ideal diamagnetism in magnetic fields smaller than the first critical field ($H < H_{c1}$), the so-called Meissner effect. However, a number of measurements carried out in superconductors has revealed a paramagnetic signal in the temperature dependence of the magnetic moment ($m(T)$) which appears upon cooling down the samples in a magnetic field through the transition temperature $T_c$ (the field cooling regime) \cite{1,2,3,4,5,6,7,8,9}. In this case, the paramagnetic contribution exceeds the diamagnetic part, leading to an overall paramagnetic signal.  This effect is called the paramagnetic  Meissner effect (PME) or Wohlleben effect \cite{1,2,3,4,5,6,7,8,9}. Three theories have been suggested to account for the origin of the effect. (i) The $d$-symmetry of the order parameter can lead to the existence of spontaneous ``paramagnetic" super-currents in the superconductor due to the presence of $\pi$-contacts \cite{10}. (ii) The Koshelev-Larkin (KL) model \cite{11} considers the redistribution of Abrikosov vortices trapped in the superconductor upon the transition to the  superconducting state, leading to the appearance of the paramagnetic signal in $m(T)$. (iii) The giant vortex state, existing in the surface superconductivity state, can also lead to a total paramagnetic signal if the temperature is decreased and the trapped flux within the giant vortex is compressed \cite{12,13}. The observations of PME in conventional superconductors with $s$-symmetry \cite{5,6,7,8,9} has indicated that $d$-symmetry of the order parameter is not a necessary condition for the appearance of PME. The giant vortex approach outlined in Refs.~\onlinecite{12,13} assumes a special geometry with the sample's extended surfaces oriented parallel to the field in order to facilitate the appearance of the giant vortex state. However, a large number of experiments have been carried out on films in fields perpendicular to the largest film surfaces. Therefore, the quantitative comparison of theoretical dependences obtained in the framework of the giant vortex model \cite{12,13} with experimental results is inadequate due to the necessity of taking into account the real boundary conditions. Nevertheless, this model can be the most appropriate for the explanation of PME in the vicinity of the superconducting transition, since fluctuations of the order parameter do not permit the formation of a pure Abrikosov vortex state, which is a necessary condition for the explanation of PME in the framework of the KL model. Well below $T_c$ (or below the irreversibility line $T_{\rm irr}$) the KL model is the most suitable one for the description of $m(T)$ in thin films. Therefore, the analysis of the experimental results obtained in this work will be carried out within the framework of this model.

In this work temperature dependences of the magnetic moment have been measured over a wide magnetic field range in YBa$_2$Cu$_3$O$_{7 - \delta}$ (YBCO) thin films grown by different methods. The experimental results obtained have been analyzed in the framework of the KL model \cite{11}, which turns out to be the most appropriate for our experimental conditions.

\section{Growth and characteristics of epitaxial YBCO films} 

Single crystalline high temperature superconducting (HTS) YBCO thin films have been investigated in this work. The films were grown by (i) pulsed-laser deposition (PLD) \cite{14} and by (ii) off-axis DC magnetron sputtering \cite{15} techniques. These techniques produce epitaxial films with the crystallographic $c$-axis oriented perpendicular to the film surface \cite{16,17}. In this case, as shown by high resolution electron microscopy \cite{streif,penn,svech}, the films grown on a mismatched substrate develop numerous out-of-plane edge dislocations. The dislocations are usually arranged in so-called dislocation walls (rows), forming 30 to 250~nm large domains. The domains are typically misaligned by $\sim 0.5^{\circ}$ to $2^{\circ}$, depending on film growth conditions. In this work, a $\sim 300$~nm thick PLD film (PP17) with $T_c \simeq 87.9$~K, as well as two $\sim 300$~nm thick magnetron sputtered films (K21 and K1509) with $T_c \simeq 86.5$~K have been investigated. The most extensively investigated K1509-film was sputtered onto a rotating sapphire substrate buffered by CeO$_2$ with a diameter of 51~mm. All the films investigated have a quite narrow transition width of $\Delta T_c = 0.2$ to 0.5~K and a high critical current density of $J_c \sim 2 \times 10^6$~A/cm$^2$ at $T = 77$~K in self-field. The film growth rate of the magnetron sputtering was approximately 0.01 to 0.02~nm/s. In these films, the layer-by-layer (nearly two-dimensional) growth mechanism is realized. In contrast, in the films grown by the PLD procedure with a growth rate of 0.1 to 0.2~nm/s, the three-dimensional island-like mechanism is most likely to occur. Therefore, the magnetron sputtered films usually have a significantly smaller density of stacking faults and accompanying dislocation loops ($\sim 10^9$~lines/cm$^2$ as estimated by high resolution electron microscopy \cite{21}) than is in the case of the PLD films. The size of the domains in the magnetron sputtered films is usually larger (up to 250~nm), as well as more ordered and equidistantly spaced than in the PLD films. The misalignment angles are typically $< 1^{\circ}$ \cite{ieee,21}. The average density of the edge dislocations is likely to be slightly smaller than in the PLD films. Similar characteristics should be expected for the K1509-film with $J_c(H = 0, T = 77 \, {\rm K}) \simeq 2.34 \times 10^6$~A/cm$^2$.

All the measured pieces of the films were of a similar rectangular shape with dimensions $\simeq 1.5$~mm$^2$.

\section{Experimental details}

The temperature and field dependences of the magnetic moment were investigated by employing a Quantum Design MPMS SQUID magnetometer in fields $|H| \le 5$~T and temperatures $5 \le T \le 95$~K. The temperature dependences were measured in the field-cooled (FC) regime, i.e. the films were cooled through $T_c$ with a magnetic field applied.
 
The MPMS SQUID detection system comprises SQUID sensing loops configured as a highly balanced second-derivative pickup coil set with a total length of $\sim 3$~cm. The coils are designed to reject the uniform field from the superconducting magnet to a precision of approximately 0.1\%. The magnetic moment of a sample is calculated from the response curve of the SQUID pickup coils which is measured as the sample moves through the coils along a scan length (typically 4~cm). The temperature of the sample is measured, depending on the temperature range, either by a sensor fixed at the null point of the pickup coils in the cooling annulus around the sample space or by a sensor located under the bottom of the sample tube \cite{MPMS}. This quite ``remote" temperature sensing is expected to be insensitive to any possible temperature gradient along the scan length. Experimentally, we did observe some temperature destabilization at scan lengths $\ge 7$~cm, which is a common feature for this kind of instrument. This observation can imply that we do indeed deal with a temperature gradient. Accordingly, the smaller the scan length is, the smaller the difference between the minimal and maximal temperatures will be. Therefore, the possibility of significant temperature fluctuations in the measured samples becomes negligible for sufficiently small scan lengths.

\begin{figure}[b]
%\vspace{-0.5cm}
\includegraphics[scale=0.34]{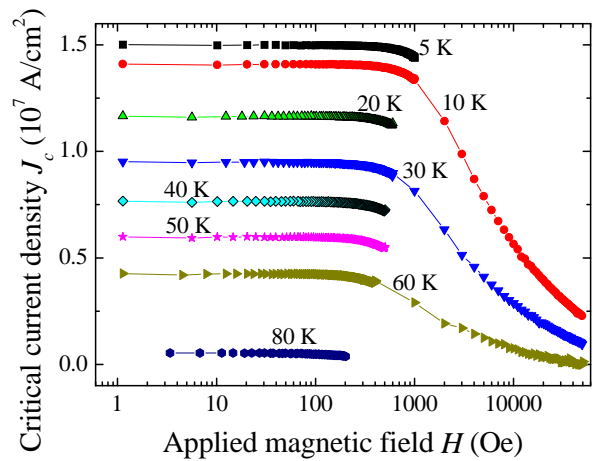}
\caption{\label{jc}Critical current density as a function of the applied magnetic field measured at different temperatures for the K1509 film.}
\end{figure}

The $J_c(H, T)$ dependencies shown in Fig.~\ref{jc}, which are necessary for the quantitative result analysis, have been obtained from the width of the magnetization loops \cite{ieee}. The so-obtained $J_c(H, T)$ were highly consistent with $J_c(H, T)$ obtained by the direct transport method and from Clem-Sanchez analysis \cite{22} of AC susceptibility measurements of the films in perpendicular fields \cite{ieee}. The $J_c(H, T)$ behavior has suggested \cite{ieee} that the mechanisms of critical current limitation can be attributed to strong pinning of vortices on linear defects, most likely edge dislocations which are perpendicular to the film surface. 

\section{Experimental results}

\begin{figure}
%\vspace{-0.5cm}
\includegraphics[scale=0.34]{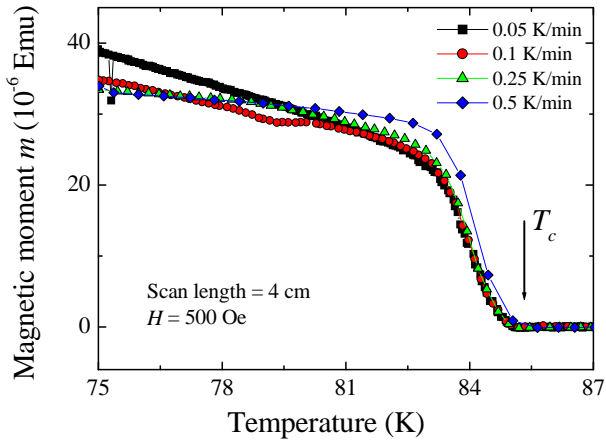}
\caption{\label{two} Temperature dependence of the FC magnetic moment in the K1509 film for different cooling rates (0.05, 0.1, 0.25 and 0.5~K/min).}
\end{figure}

\begin{figure}
%\vspace{-0.5cm}
\includegraphics[scale=0.34]{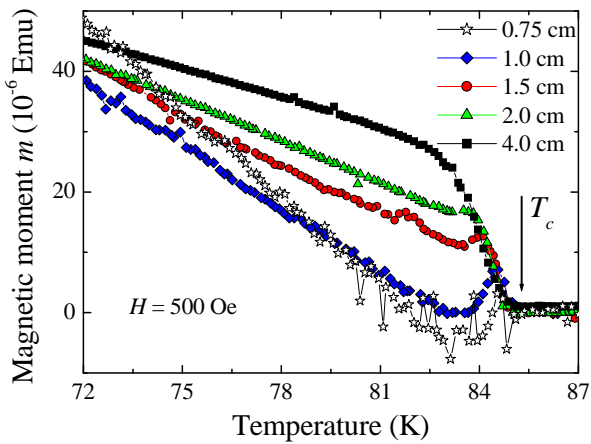}
\caption{\label{three}Temperature dependence of the FC magnetic moment in the K1509 film for different scan lengths (4, 2, 1.5, 1 and 0.75~cm) and 0.1~K/min cooling rate.}
\end{figure}

\begin{figure}
%\vspace{-0.5cm}
\includegraphics[scale=0.34]{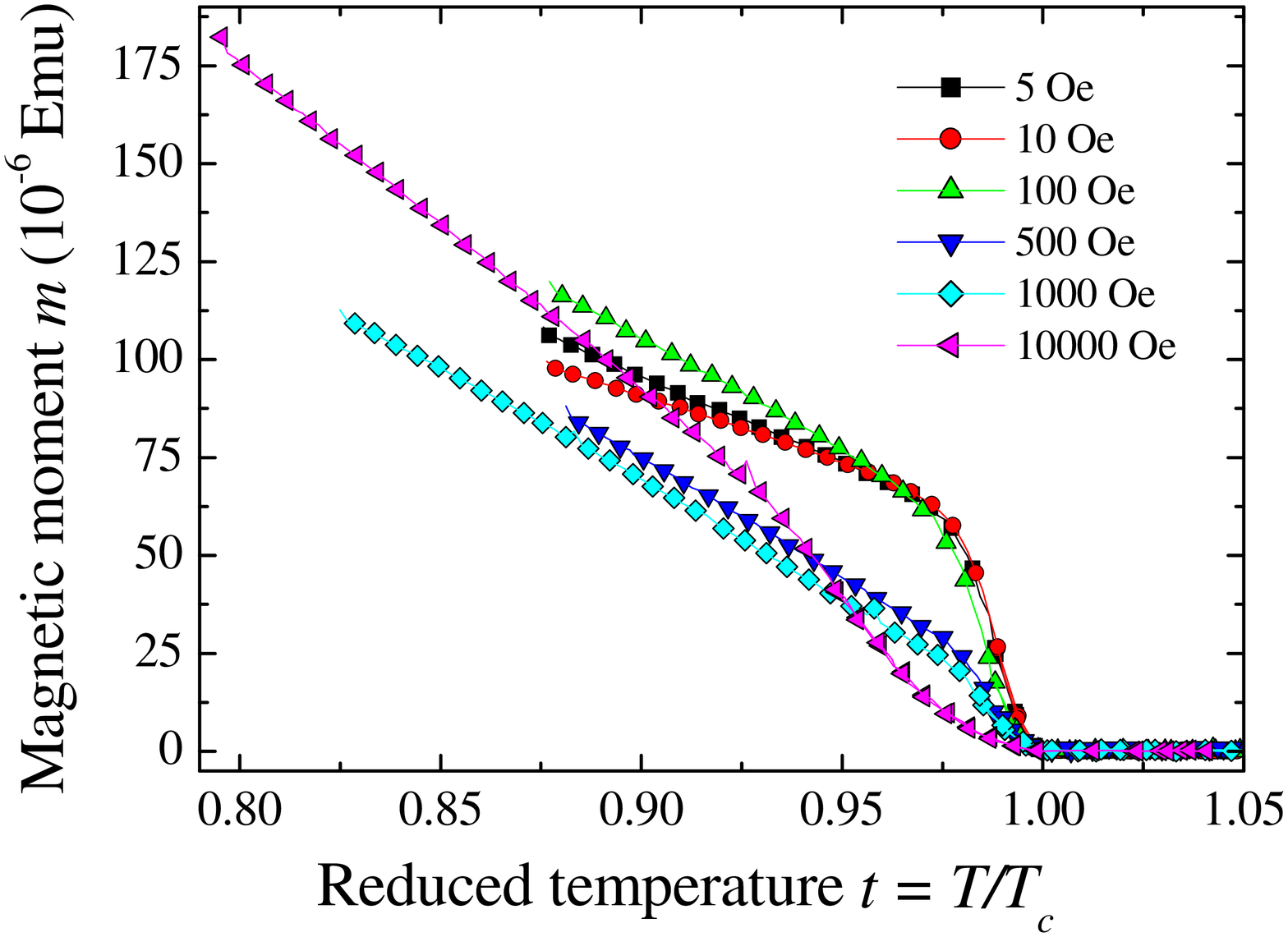}
\caption{\label{four}The field-cooled magnetic moment as a function of the reduced temperature $t = T/T_c(H)$ in the K1509 film measured with 0.1~K/min cooling rate at different magnetic fields applied perpendicular to the film.}
\end{figure}

The FC dependences of $m(T)$ measured in the K1509, PP17 and K21 films are qualitatively consistent with each other, and therefore, in what follows, the results for the K1509-film will be shown. The measurements at different cooling rates (0.05, 0.1, 0.25 and 0.5~K/min) and different scan lengths (4, 2, 1.5, 1 and 0.75~cm) are exhibited in Figs.~\ref{two} and \ref{three}, respectively. As can be seen, different cooling rates do not have a significant qualitative influence on the $m(T)$ behavior. The small discrepancies might be explained either by temperature lags at the highest sweep rates or by the influence of the magnetic flux distribution effects described in the next section. In contrast, changes in scan length qualitatively modify the $m$ behavior below $T_c$. As the temperature is further decreased the $m$ signal rises, and the signals become comparable for all the measured scan lengths. The $m(T)$ behavior has also been measured at different magnetic fields (Fig.~\ref{four}). As the field changed by more than three orders of magnitude the $m$ value was changed by less than one order of magnitude over the temperature range of the measurements.

The characteristic features of the dependencies in Figs.~\ref{two}-\ref{four} are as follows. (i) The diamagnetic response is absent below $T_c$ at the scan length $\ge 1$~cm. At shorter scan lengths a magnetic moment $< 0$ can be measured in the vicinity of $T_c$. (ii) The change in $m$ is rather moderate compared to the large change in $H$. In addition, $m(T)$ depends non-monotonously on the applied field. (iii) The paramagnetic value of $m$ monotonically increases in decreasing temperature, indicating the absence of saturation. Feature (iii) has been obtained in a number of works \cite{1,4}, whereas features (i) and (ii) are observed for the first time.

The measurements indicate that the profile of the trapped magnetic flux which determines the value of the $m$ signal may be influenced by certain experimental conditions, such as the scan length and the applied magnetic field. As will be shown in the next section, changes in the $m$ value observed in the experiments are governed by relatively small changes in the vortex compression. Moreover, we show that the vortex compression in the films is {\em extremely} sensitive to seemingly negligible changes in experimental conditions, which can explain various experimental results reported on the paramagnetic Meissner effect in the literature and in Fig.~\ref{three}.

\section{Theoretical consideration}

\begin{figure}
\hspace{-1.cm}
\includegraphics[scale=1]{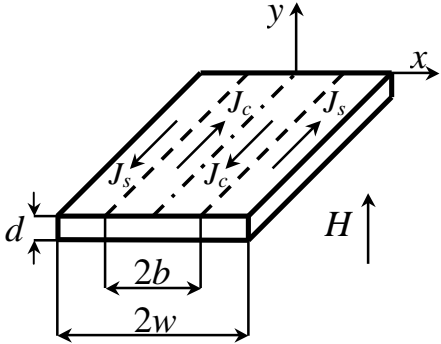}
\vspace{-14.5cm}
\caption{\label{model}The complete Bean state is schematically shown in a superconducting film upon cooling in a perpendicular field (the model of KL \cite{11}). The vortices concentrate in the $|x| < |b| < |w|$ region, whereas the Meissner shielding currents ($J_s$) flows in the $|b| < |x| < |w|$ region.}
\end{figure}

As was mentioned in Sect.~\ref{int}, from the point of view of the real geometry of the experiment, the most likely mechanism for the appearance of the PME is the inhomogeneous redistribution of vortices upon cooling a YBCO film below $T_c$. Note that in this case the term -- Paramagnetic {\em Meissner} Effect (PME) - would inappropriately reflect the nature of the effect, which should be rather referred to as the Paramagnetic Effect or the Positive Magnetization Effect (PME). Vortex distributions leading to the PME are considered in the model developed by Koshelev and Larkin \cite{11}. The geometry of this model is shown in Fig.~\ref{model}. The model assumes that if the film is cooled down below $T_c$ in a perpendicular field $H_z$ then, firstly, the film enters the vortex (mixed) state and, secondly, the gradient of vortex density in the region $|x| < |b| < |w|$ corresponds to $J_c$. Such a density can be formed due to the interaction between vortices and surface Meissner currents (``sheet currents"), which forces vortices inward from the film edges. $b$ characterizes the degree of the vortex compression and unambiguously defines the magnetic flux through the film:
\begin{equation}
f = \Phi_H - \frac{4 d J_c}{c H_z} \Phi_J \, ,
\label{e1} \end{equation}
provided that $J(|x| < |b|) = J_c$ and $f = \Phi_{\Sigma}/2wH_z$. $\Phi_{\Sigma}$ is the full magnetic flux through the film, $\Phi_H = E(k) - (1-k^2)K(k)$ is the field contribution, and $\Phi_J = [E(k) - (1-k^2)K(k)] \ln[(1+k)/(1-k)] - 2kK(k)$ is the shielding current contribution. $E(k)$ and $K(k)$ are the complete elliptic integrals of the first and second kinds, respectively, $k = b/w$.

In the case of a $2w$ wide, $d$ thick and infinitely long film (Fig.~\ref{model}), the density of the magnetic moment $m_{\rho}(H_z, T)$ is the sum of two components \cite{11}: 
\begin{equation}
m_{\rho} = - \frac{H_z w}{8d} \left ( m_H - \frac{4 d J_c}{c H_z} m_J \right ) \, ,
\label{e2} \end{equation}
where $m_H = 1 - k^2$ is the diamagnetic part of the Meissner currents and  $m_J = (1 - k^2) \ln[(1+k)/(1-k)] + 2k$ is the paramagnetic part, arising due to the gradient of the vortex density. The total magnetic moment $m(H_z, T)$ is $m_{\rho}(H_z, T)$ multiplied by the film volume.

The $J_c$ dependence on temperature results in $b$ being temperature dependent. In turn, this leads to two independent variables, $J_c$ and $k$, in Eq.~(\ref{e2}). This fact does not allow us to directly calculate the critical current density and parameter $k$ (and $f$) from $m(H_z, T)$ alone. Independent measurements on these films provided us with $J_c(H_z, T)$ \cite{18,19}, which turned out to be well described by the following empirical formula 
\begin{equation}
J_c(H_z,T) = J_{c0}(H_z)(1-T/T_c)^{\alpha} \, ,
\label{e3} \end{equation}
where $J_{c0}(H_z) = J_{c0}(0) \widetilde{\alpha} \ln(H^{\ast}/H_z)$ at $H_z \ge H_{tr}$ and $J_{c0}(H_z) = J_{c0}(0)$ at $H_z \le H_{tr}$. $H_{tr} = H^{\ast} \exp(-1/\widetilde{\alpha})$ defines the crossover field from the plateau to the decreasing part of $J_c(H_z)$ (Fig.~\ref{jc}) \cite{18,19}. The temperature dependence of $H^{\ast}$ is $H^{\ast} \simeq H_{\rm eff} (1-T/T_c)$. For the YBCO films the parameters $J_{c0}(0)$, $\widetilde{\alpha}$, $\alpha$ and $H_{\rm eff}$ are determined by the conditions and method of preparation. The variation of these parameters is negligible for films obtained under the same preparation conditions. These parameters can be independently obtained from transport, AC susceptibility and magnetization measurements at low temperatures. For the K1509 film, the following values were obtained $\alpha \simeq 1.13$, $\widetilde{\alpha} \simeq 0.24$ and $H_{\rm eff} \simeq 6.4 \times 10^4$~Oe \cite{ieee}. $J_{c0}(0)$ was chosen so that $J_c$ at $T = 77$~K is equal to the experimentally obtained value of $2.34 \times 10^6$~A/cm$^2$, therefore, $J_{c0}(0) = 2.34 \times 10^6 (1- 77/T_c)^{-\alpha}$~A/cm$^2$.

\begin{figure}
%\vspace{-0.5cm}
\includegraphics[scale=0.34]{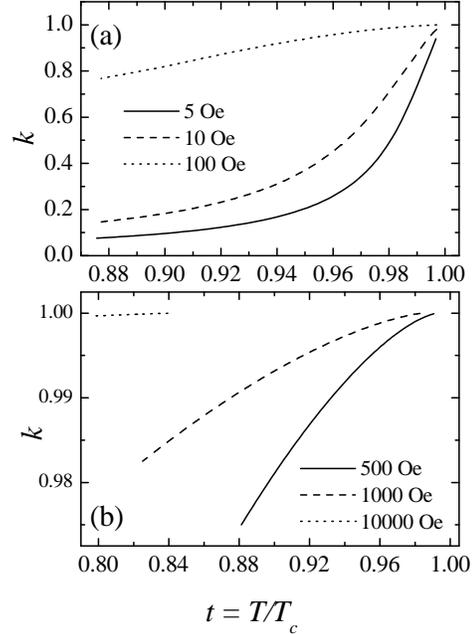}
\caption{\label{f5}Temperature dependence of the vortex compression coefficient $k = b/w$ for small (a) and large external fields(b).}
\end{figure}

\begin{figure}
%\vspace{-0.5cm}
\includegraphics[scale=0.34]{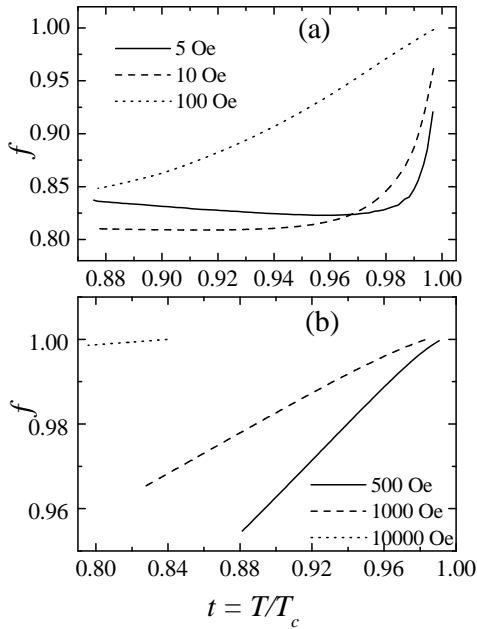}
\caption{\label{f6}Temperature dependence of the magnetic flux in the FC state below $T_c$, normalized to the flux in the film in its normal state for small (a) and large fields(b).}
\end{figure}

For the detailed quantitative analysis of the experimental data, the dependencies obtained for the K1509 film (Fig.~\ref{four}) have been used.  Using Eqs.~(\ref{e1}) and (\ref{e2}), as well as the experimentally obtained formula~(\ref{e3}), we have calculated the temperature dependences of the parameters $k$ and $f$ for the vortex arrangement in the film (Figs.~\ref{f5} and \ref{f6}, respectively). In the immediate vicinity of $T_c$, where the critical current density is rather small, the PME cannot be obtained in the frame of the KL model \cite{11}. Therefore, the behavior in Figs.~\ref{f5} and \ref{f6} was obtained within the temperature range of the KL model applicability. Note that our experimental geometry and, correspondingly, our current distribution are different from the geometry shown in Fig.~\ref{model}. However, there are no analytical expressions for $m(k, H, J_c)$ for our case of finite sample length. The existing expressions for a circular sample \cite{11} would only insignificantly shift the $f(t)$, $k(t)$ curves along the corresponding ordinate axis, where $t = T/T_c(H)$ with $T_c(H)$ being the critical temperature obtained from the experiments performed at corresponding fields. Moreover, the magnetic response of a sample weakly depends on its shape \cite{wurl}. Thus, the original KL model can be employed.

It is shown that for relatively high fields ($H_z > 100$~Oe) the $m(T)$ behavior corresponds to the case of weak vortex compression ($1-k << 1$). For low fields ($< 10$~Oe) the compression becomes quite large ($k << 1$) at low temperatures, which does not allow us to use the asymptotical relationship between the parameters $m$, $k$ and $f$ obtained in Ref.~\onlinecite{11}, justifying the use of the exact expressions~(\ref{e1}) and (\ref{e2}).

In the low field region a small difference in the paramagnetic moment (Fig.~\ref{four}) can be entirely explained by a small change in the flux and the compression of the vortex arrangement in the film (for example, as a result of flux creep). This is due to the fact that over this field range $J_c$ does not depend on the applied field (Fig.~\ref{jc}). In high fields, the degradation of $J_c$ in increasing field is compensated by an increase in the magnetic flux in the film (Fig.~\ref{f6}) and a decrease in the vortex compression (Fig.~\ref{f5}). In this case, the weak compression regime ($1-k << 1$) is realized, i.e. the vortices almost entirely fill up the film. As the temperature decreases, vortices near the film's edges can freely leave the film (due to flux creep) via edge micro-cracks and other defects inherent to every film \cite{zeldov}. This process results in increasing compression and decreasing flux in decreasing temperature.

Despite the fact that the $m(t)$ dependence on the field is non-monotonic (Fig.~\ref{four}), $f(t)$ and $k(t)$ monotonically increase with increasing field for all the fields except $H = 5$~Oe (Fig.~\ref{f6}a). Moreover, in contrast to the $f(t)$ behavior for all the other measured fields, $f(t)$ at $H = 5$~Oe starts increasing in decreasing temperature at $t < 0.965$. This deviation of $f(t)$ from monotonic behavior as a function of field and temperature at the lowest measurement field can be explained by the entry of additional vortices from the film edges as the temperature is decreased. The conditions for entry are defined by the sheet currents, which, in turn, significantly depend on the degree of vortex compression in the film. Vortex entry through the edges of the film in this case is facilitated by the small number of vortices and the strong compression in the field as a result of relatively large sheet currents.

\subsection{$m(T)$ behavior as a function of the scan length}

\begin{figure}
%\vspace{-0.5cm}
\includegraphics[scale=0.34]{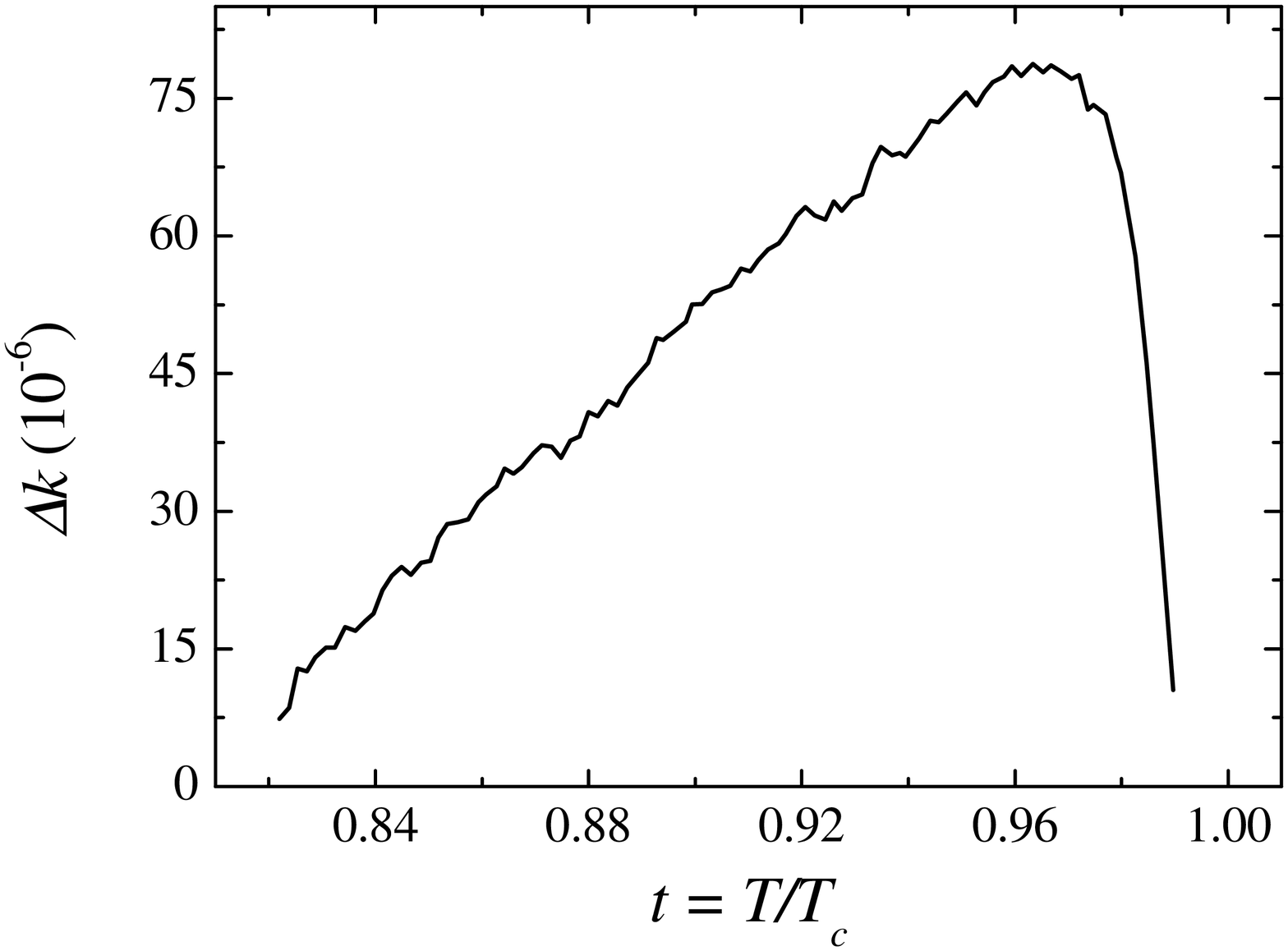}
\caption{\label{field}Temperature dependence of the relative change of $\Delta k = k(4 \, {\rm cm}) - k(1 \, {\rm cm})$ caused by the change of the scan length from 4~cm to 1~cm at $H = 500$~Oe.}
\end{figure}

The parameters of the KL theory, the vortex compression $k$ and the magnetic flux $f$ within the film also enable us to quantitatively account for the $m(T)$ behavior as a function of the scan length (Fig.~\ref{three}). Indeed, the PME can be affected by inhomogeneity of the magnetic field inside the magnet \cite{scan}, which can influence the flux profile within the film. From the curves in Fig.~\ref{three} it is possible to estimate a magnetic field change ($\Delta H$) or field gradient in the magnet, which would result in the different $m(T)$ behavior as a function of the scan length. In the general case, the magnetic moment can be written as $m = m(k,H)$, so its change is 
\begin{equation}
\Delta m \simeq \frac{\partial m}{\partial k} \Delta k + \frac{\partial m}{\partial H} \Delta H \, .
\label{m1} \end{equation}
It can be equivalently rewritten as
\begin{equation}
\Delta m \simeq \left ( \frac{\partial m}{\partial k} \frac{\partial k}{\partial H} + \frac{\partial m}{\partial H} \right ) \Delta H \, ,
\label{m2} \end{equation}
where $\partial k/\partial H$ is unknown function and can only be roughly estimated with the help of the experimental data in Fig.~\ref{f5}. Therefore, we have used Eq.~(\ref{m1}) for $\Delta H$ and $\Delta k$ estimations.

Assuming $\Delta H = 0$, in Fig.~\ref{field} we show the calculated change of the compression coefficient ($\Delta k = k(4 \, {\rm cm}) - k(1 \, {\rm cm})$) of the frozen vortex ensemble in the K1509 film measured under the same conditions, but with two different scan lengths: 4~cm and 1~cm (Fig.~\ref{three}). As can be seen, the change in the maximum does not exceed 0.01\%, which corresponds to a maximum change in the magnetic flux $\Delta f$ within the film of $\simeq 0.03$\%. The dependence of $\Delta f = f(4 \, {\rm cm}) - f(1 \, {\rm cm})$ can be calculated either similarly to $\Delta k$ by direct solution of Eqs.~(\ref{e1}-\ref{e3}), or in agreement with the expression $\Delta f (t, J_c(t)) = (\partial f(t)/ \partial k) \Delta k(t)$, where  $\Delta k(t)$ is the dependence in Fig.~\ref{field}. Therefore,  $\Delta f (t)$ behaves essentially the same as $\Delta k(t)$. 

If we now assume that in Eq.~(\ref{m1}) $\Delta k = 0$ (correspondingly $\Delta f = 0$), which is {\em a priori} not correct, we can calculate the {\it maximal} field change ($\Delta H^{\rm max}$) responsible for the {\em maximal} $\Delta m = m(4~{\rm cm}) - m(1~{\rm cm})$: $\Delta H^{\rm max} \simeq 50$~Oe for $H = 500$~Oe (Fig.~\ref{three}). Apparently, $\Delta H^{\rm max} < 0$, since $\partial m/ \partial H < 0$ and $\Delta m > 0$ (see Eqs.~(\ref{e2}), (\ref{m1}), and Fig.~\ref{three}). Since $\Delta k > 0$, this field change can be considered as a {\em significantly overestimated} 10\% field gradient over the 4~cm scan length in the magnet bore along its axis.

It is worth mentioning that the magnetic flux change $\Delta f$ within the film does not necessarily imply the entry or exit of vortices. This is because for a given amount of vortices the flux is defined by their gradient ($J_c$) and compression ($k$). This means that with a fixed amount of vortices we get different values of $f$ by changing $k$. Changes in $k$ mainly occur due to changes in the sheet currents, flowing in vortex free film regions (Fig.~\ref{model}). Thus, the discrepancies between the $m(T)$ curves in Fig.~\ref{three} are due to changes in the magnetic flux profile in the film, which are, in turn, caused by inhomogeneous field distribution in the magnet.

Moreover, as the sample moves through the SQUID pickup coils it is likely to experience not only the field inhomogeneity, but also the temperature gradient along the length of the scan. The latter factor would play a boosting role in redistribution of the flux profile, which would occur due to temperature variations within the sample. In the vicinity of $T_c$, the effect of the temperature variations can be even more pronounced due to strong fluctuations of the superconducting order parameter. The influence of the temperature gradient is a rather unpredictable factor, which may depend on temperature, the parameters of the temperature controller and the level of liquid helium inside the cryostat. In addition, the field gradient of a magnet can also be a variable. It may depend on the field pre-history and the remnant field \cite{scan}. Therefore, the influence of these factors on the flux profile may have led to somewhat non-reproducible $m(T)$ behavior in Fig.~\ref{three} and might explain various experimental results available in the literature \cite{1,3,5,7,8,9}.

The question arises why three orders of magnitude of the external field introduce a relatively small change in $m(T)$ behavior (Fig.~\ref{four}), whereas a small field gradient $< 10$\% can introduce not only comparative changes in $m(T)$, but also a qualitative change at small scan lengths (Fig.~\ref{three}). The first part of the question is dealt with in the previous subsection, where the compensating interplay between $k(T,H)$ and $J_c(T,H)$ is identified in the quasi-static, steady scenario. What if the field is suddenly changed (due to the gradient). This change would affect the sheet currents and, correspondingly, the vortex arrangement, i.e. $k$. The degree of the vortex re-arrangement depends on the vortex mobility, which, in turn, can depend on $H$ and $T$. On the other hand, the $J_c(H)$ change would be negligible, since $J_c \propto \log H$ (Eq.~(\ref{e3})). Therefore, the change in $k$ is not compensated by the degradation of $J_c$, as discussed in the previous subsection. This imbalance between the paramagnetic and diamagnetic contributions can lead to considerable, even qualitative changes in the $m(T)$ behavior observed  (Fig.~\ref{three}). 

An additional factor, which can contribute to differences in FC $m(T)$ behaviors for similar samples, is structural imperfection. In films as in our case, differences in the quality of the film edges can also define the structure of the vortex ensemble in the films \cite{zeldov} and, together with field and temperature gradients, lead to some differences between $m(T)$ behaviors observed for different films produced and measured under similar conditions.

\section{Conclusion}
 
In conclusion, the origin of the paramagnetic Meissner effect in experimental $m(H, T)$ dependences obtained in YBCO films was shown to be in good quantitative agreement with the model of vortex compression proposed in Ref.~\onlinecite{11}. For a consistent description of the $m(H, T)$ dependencies in the frame of the model, the experimental $J_c(T, H)$ dependence has to be taken into account, and this was independently determined in Refs.~\onlinecite{ieee,18,19}. Upon changing the applied field by three orders of magnitude, the absolute value of the PME is changed by less than one order of magnitude. This change can be explained by, firstly, the field independent behavior of $J_c$ at $H < 100$~Oe and, secondly, by the degradation of $J_c(H)$ at $H > 100$~Oe, which is compensated by a sharp decrease of the vortex compression ($1-k <<1$) and magnetic flux in the film. The discrepancies between the experimental curves of $m(T)$ obtained at different scan lengths can be accounted for by the influence of field and temperature gradients along the scan length of the samples.

\begin{acknowledgments}
We thank A. L. Kasatkin, A. V. Semenov and J. Horvat for useful and fulfilling discussions and T. Silver for reading the manuscript and critical remarks. D.A.L. and V.M.P. are grateful to the University of Leipzig for support and hospitality, as well as to Deutsche Akademischer Austausch Dienst (DAAD) for financial support through Leonard Euler's Programme. This work was also financially supported by the Australian Research Council.
\end{acknowledgments}

%\bibliography{bib}

\begin{thebibliography}{99}

\bibitem{1}F. T. Dias, P. Pureur, P. Rodrigues Jr., and X. Obradors, Physica C {\bf 354}, 219 (2001).
\bibitem{2}S. Riedling, G. Br\"auchle, R. Lucht, K. R\"ohberg, H. V. L\"ohneysen, H. Claus, A. Erb, and G. M\"uller-Vogt, Phys. Rev. B {\bf 49}, 13283 (1994).
\bibitem{3}R. Lucht, H. V. L\"ohneysen, H. Claus, M. Kl\"aser, and G. M\"uller-Vogt, Phys. Rev. B {\bf 52}, 9724 (1995).
\bibitem{4}A. I. Rykov, S. Tajima, and F. V. Kusmartsev, Phys. Rev. B {\bf 55}, 8557 (1997).
\bibitem{5}L. Pust, L. E. Wenger, and M. R. Koblischka, Phys. Rev. B {\bf 58}, 14191 (1998).
\bibitem{6}A. Terentiev, D.B. Watkins, L. E. De Long, D.J. Morgan, and J.B. Ketterson, Phys. Rev. B {\bf 60}, R761 (1999).
\bibitem{7}P. Kostic, B.Veal, A. P. Paulikas, U. Welp, V. R. Todt, C. Gu, U. Geiser, J. M. Williams, K. D. Carlson, and R. A. Klemm, Phys. Rev. B {\bf 53}, 791 (1996).
\bibitem{8}T. M. Rice and M. Sigrist, Phys. Rev. B {\bf 55}, 14647 (1997).
\bibitem{9}P. Kostic, B. Veal, A. P. Paulikas, U. Welp, V. R. Todt, C. Gu, U. Geiser, J. M. Williams, K. D. Carlson, and R. A. Klemm, Phys. Rev. B {\bf 55}, 14649 (1997).
\bibitem{10}M. Sigrist and T. M. Rice, J. Phys. Soc. Jpn. {\bf 61}, 4283 (1992).
\bibitem{11}A. E. Koshelev and A. I. Larkin, Phys. Rev. B {\bf 52}, 13559 (1995).
\bibitem{12}V. V. Moshchalkov, X. G. Qiu, and V. Bruyndoncx, Phys. Rev. B {\bf 55}, 11793 (1997).
\bibitem{13}G. F. Zharkov, Phys. Rev. B {\bf 63}, 214502 (2001).
\bibitem{14}I. Peshko, V. Flis, V. Matsui, J. Phys. D: Appl. Phys. {\bf 34}, 732 (2001).
\bibitem{15}V. A. Komashko, A. G. Popov, V. L. Svetchnikov, A. V. Pronin, V. S. Melnikov, A. Yu. Galkin, V. M. Pan, C. L. Snead, M. Suenaga,  Supercond. Sci. Technol. {\bf 13}, 209 (2000).
\bibitem{16}V. M. Pan, V. S. Flis, O. P. Karasevska, V. I. Matsui, I. I. Peshko, V. L. Svetchnikov, M. Lorenz, A. N. Ivanyuta, G. A. Melkov, E. A. Pashitskii, H. W. Zandbergen, J. Superconductivity: Incorporating Novel Magn. {\bf 14}, 105 (2001).
\bibitem{17}V. M. Pan, V. S. Flis, V. A. Komashko, O. P. Karasevska, V.L. Svetchnikov, M. Lorenz, A. N. Ivanyuta, G. A. Melkov, E. A. Pashitskii, H. W. Zandbergen, IEEE Trans. Appl. Supercond. {\bf 11}, 3960 (2001).
\bibitem{streif}S. K. Streiffer, B. M. Lairson, C. B. Eom, B. M. Clemens, J. C. Bravman, and T. H. Geballe, Phys. Rev. B {\bf 43}, 13007 (1991).
\bibitem{penn}S. J. Pennycook, M. F. Chisholm, D. E. Jesson, R. Feenstre, S. Zhu, X. Y. Zheng, and D. J. Lowndes, Physica C {\bf 202}, 1 (1992).
\bibitem{svech}V. L. Svetchnikov, V. M. Pan, Ch. Traeholt, and H. W. Zandbergen, IEEE Trans. Appl. Supercond. {\bf 7}, 1396 (1997).
\bibitem{21}V. M. Pan, C. G. Tretiatchenko, V. S. Flis, V. A. Komashko, E. A. Pashitskii, A. N. Ivanyuta, G. A. Melkov, H. W. Zandbergen, and V. L. Svetchnikov, Journal of Superconductivity: Incorporating Novel Magnetism (2003), in press.
\bibitem{ieee}V. M. Pan, E. A. Pashitskii, S. M. Ryabchenko, V. A. Komashko, A. V. Pan, S. X. Dou, A. L. Kasatkin, A. V. Semenov, C. G. Tretiatchenko, and Y. V. Fedotov, IEEE Trans. Appl. Supercond. {\bf  13}, June (2003), in press.
\bibitem{MPMS}{\em MPMS XL Operation Manual}, Quantum Design, San Diego, CA, 1999.

\bibitem{22}J. R. Clem and A. Sanchez, Phys. Rev. B {\bf 50}, 9355 (1994).
%\bibitem{23}E. Zeldov, A. I. Larkin, V. B. Geshkenbein, M. Konczykowski, D. %Majer, B. Khaykovich, V. M. Vinokur and H. Shtrikman, Phys. Rev. Lett. {\bf %73}, 1428 (1994).
\bibitem{wurl}M. Wurlitzer, M. Lorenz, K. Zimmer, and P. Esquinazi, Phys. Rev. B {\bf 55}, 11816 (1997).
\bibitem{18}Yu. V. Fedotov, S. M. Ryabchenko, E. A. Pashitskii, A. V. Semenov, V. I. Vakaryuk, V. S. Flis, and V. M. Pan, Physica C {\bf 372-376}, 1091 (2002).
\bibitem{19}Yu. V. Fedotov, S. M. Ryabchenko, E. A. Pashitskii, A. V. Semenov, V. I.  Vakaryuk, V. M. Pan, and V. S. Flis, Low Temp. Phys. {\bf 28}, 172 (2002).
\bibitem{zeldov}Y. Paltiel, E. Zeldov, Y. N. Myasoedov, H. Shtrikman, S. Bhattacharya, M. J. Higgins, Z. L Xiao, E. Y. Andrei, P. L. Gammel, and D. J. Bishop, Nature {\bf 403}, 398 (2000).
\bibitem{scan}F. J. Blunt, A. R. Perry, A. M. Campbell, and R. S. Liu, Physica C {\bf 175}, 539 (1991).

\end{thebibliography}

\end{document}